# From 600 Tools to 1 Console: A UX-Driven Transformation

Mariann Kornelia Smith
Google
Seattle, Washington, USA
mariannksmith@google.com

Jacqueline Meijer-Irons
Google
Seattle, Washington, USA
jjmeijer@google.com

Andrew Millar
Google
London, UK
andrewmillar@google.com

## ABSTRACT

In 2021 the Technical Infrastructure (TI) User Experience (UX) team sent a survey to 10,000 Google Developers (Googlers) and uncovered that Google's internal infrastructure tools were fragmented and inefficient, hindering developers' productivity. Using user centered research and design methodologies the team first created a story map and service blueprint to visualize the relationship between internal applications, then formulated a strategic vision to consolidate tools, streamline workflows, and measure the impact of their work. We secured executive buy-in and delivered incremental improvements.

I would love to join the UXR POV workshop at CHI and transform some of the research program we used into plays, as well as learn from other teams and leverage new plays in some areas (eg. measuring success).

## 1 INTRODUCTION

"Figuring out why a Flume temp file appears to get deleted... is effectively impossible. It's probably a quota issue, or a scratch issue, or a colossus issue, or a DAX issue, or a CellScorer issue, or a logs-cell issue, rather than a Flume issue, but debugging the intersection of all of these factors is effectively impossible." - Technical Infrastructure (TI) Customer Survey, 2021

It is important for Google to be able to prioritize computing efficiency and rapid feature development within its internal applications. While this has driven innovation, it has also created usability challenges for Googlers, impacting their productivity and leading to excess time spent on specific tasks. Our team identified two main sources of these challenges.

The first is fragmented information architecture. There are over 1000 distinct, internal tools that were developed by various teams at Google and are at different levels of support ranging from fully supported (managed by the team that developed the solution) to unsupported and niche solutions to a problem that doesn't exist anymore. We found that even with managed services there are often documentation gaps, scattered dashboards, and inconsistent graphical user interfaces (UIs). These tools create complexity and impede efficient workflow navigation.

The second is inconsistencies. Because the internal tools were developed in small teams, they often lack the dedicated user experience (UX) focus given to external-facing Google products, resulting in solutions that slow down users. This friction is evident in direct feedback from Googlers.

Recognizing the need for improvement, Google invested in a dedicated UX team to enhance developer productivity through improved usability. This case study examines our team's journey to identify the core issue, develop a vision to improve developer workflows, establish a measurement framework to track progress, and deliver a minimum viable product (MVP) demonstrating impact.

## 2 FROM AMBIGUITY TO STRATEGIC VISION

### 2.1 The Challenge

We identified the challenge: Google's internal infrastructure tools are hard to discover, different in look and feel, hard to integrate, difficult to onboard to, and time consuming to troubleshoot.

We gained this insight from our Google-wide survey that we sent out in November 2021 to 10,000 randomly selected Googlers meeting specific criteria, such as tenure, role, frequency of tool use, self identified level of proficiency. The survey's primary objective was to establish a foundational understanding of the domain space and to measure baseline satisfaction metrics in various categories. Of the 1,300 Googlers who filled out the survey, 749 provided qualitative feedback.

Analysis of the survey revealed that debuggability was the most pressing issue accounting for 67 percent of open-ended feedback. Although the necessary information was available, users reported spending a significant amount of time locating it across disparate dashboards and UIs. This process often involved numerous clicks and extensive scrolling. Performance and reliability (a term users used interchangeably to describe similar challenges) represented another major area of dissatisfaction, comprising 19 percent of open-ended feedback. These concerns encompassed both the time required to initiate tasks and overall system optimization. Based on survey feedback we formed three primary hypotheses: Hypothesis 1: A streamlined developer experience that consolidated disparate tools would probably enhance developer productivity across TI products and services.

Hypothesis 2: A shared design language across Compute, Data Processing, Networking, and Storage would reduce cognitive load and friction, accelerate debugging processes, and enhance user satisfaction with perceived usability.

Hypothesis 3: Establishing a measurement framework was needed to show progress and impact of UX work.

To test hypotheses 1 and 2, we did a literature review[1] of both internal and external research to persuade ourselves that consolidating disparate tools and applying a consistent design language would likely improve productivity.

To test hypothesis 3 we did stakeholder interviews and a retrospective analysis of past success measurement related projects at Google. We learned that Google is littered with dashboards, trackers, and initiatives to identify metrics that matter for teams and organizations. These initiatives always have great initial interest from all levels of management but a quarter later the projects run out of steam and the dashboards get no attention. And even though we reviewed the most popular measurement frameworks we knew that building another dashboard wasn't the right solution for our team.

Based on evidence that pointed to abandoned dashboards, unfinished projects, and big up-front costs we concluded that yet another dashboard probably wouldn't have the desired impact. Instead, we identified one key metric to measure our success: amount of full-time software engineer (SWE) capacity saved in a year.

## 2.2 The Value of UX Improvements

We wanted to build a unified console to streamline developer experience and reduce the cognitive load of context switching. The console would provide users with timely and useful observability and status information about their projects, enabling the TI organization to transition toward a managed service offering, where they would take on the responsibility of managing and maintaining the infrastructure tools for developers, reducing the burden on its customers. We defined a metric to measure return on investment (ROI).

$$ROI = \frac{SWE\ hours\ saved}{SWE\ hours\ invested}$$

To estimate developer hours saved, we needed to establish a benchmark so we designed a mixed-method research program that included in-depth user interviews and logs analysis. To estimate developer hours invested, we created an estimation framework that gave TI developers a consistent framework to estimate the amount of time required to complete a project.

Then we invited 25 customer survey participants that opted into further discussions from various organizations at Google to 60 minute long, one-on-one, semi-structured interviews[4]. We completed discussions with 23 users. Our goal was to provide actionable insights and define next steps for UX, engineering, and product management teams to enhance user experience. We focused primarily on four key areas:

- To understand users' primary information needs during debugging.
- To map users' workflows.
- To visualize the ecosystem of tools used.
- To identify areas for improvement.

After the interviews we facilitated a cross-functional workshop between UX, Engineering (SWE), and Product Management (PM) partners where we used data from the interviews as well as expert knowledge from stakeholders to create a comprehensive story map that captured the end-to-end user journey. This map highlighted user goals, tasks, touch-points, and areas for further exploration.

Next we defined jobs to be done (CUJs)[<empty citation>] and validated them with users through another set of semi-structured interviews. Based on feedback from users we were able to add previously missed tasks. These journeys served as a foundation for understanding users' needs and gaps in information and tooling.

Once we validated the CUJs, we built upon the concept by developing a service blueprint[3] that mapped user interactions, pain points, and opportunities across various service layers. As figure 1 shows our blueprint helped us identify root causes of problems by mapping each pain point to a certain stage of the journey making it clear where in the process users struggled and with what. Based on this information we generated ideas for improvement.

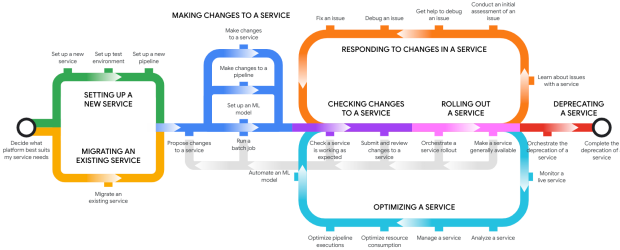

**Figure 1: The service model the team created based on information gathered in the service blueprint**

The final deliverable of the first part of our research program was a high-level service model that we created to provide a simplified overview of the end-to-end journey. It consolidated insights from the individual CUJs as well as the service blueprint stages.

Via trends analysis in feedback and quantitative user data we identified both long- and short-term opportunities to have a positive impact on focus areas and user experience with low to high effort.

Prioritizing low-effort, high-impact work resulted in improvements such as increased visibility of features, improved affordances for interactive elements (like indicating hover states), and improved consistency within the wider ecosystem of products used by Googlers.

We also addressed popular trends in feedback around improving density and readability using the rapid iterative testing and evaluation (R.I.T.E.) method that allowed us to test design concepts, collect feedback, and iterate on the proposed solution. We repeated this cycle three times before we got to a state that both the UX team and stakeholders were satisfied.

We recognized that addressing immediate concerns alone would not suffice long-term. At this time we envisioned a unified console to improve efficiency, developer productivity, and streamline Google's internal infrastructure; a new direction that eliminates redundant efforts and centralizes solutions for infrastructure needs.

"Improve company velocity, efficiency, and productivity and deliver durable cost savings" has been one of Alphabet's annual objectives. We decided to lean into efficiency and productivity. Because engineering work is estimated based on the amount of time it will take one developer to deliver the project, we defined the number of SWEs saved as the number of full-time Google developer resources saved as our metric of success. We calculated three different levels of savings that our project would mean to the company:

- Conservative estimate: 69 SWEs.
- Moderate estimate: 208 SWEs.
- Aggressive estimate: 1,663 SWEs.

The three different levels of savings depended on initial resource investment and the state of the delivered solution at launch time. So while the conservative estimate only projected to save 69 SWEs a year, it was the cheapest solution. And while the aggressive estimate promised 1,663 SWEs saved per year it also had a much higher upfront cost.

## 2.3 Strategic Vision

During Google Cloud's innovation week (a four-day long design sprint that occurs twice a year with the goal of creating space to explore new ideas and push the boundaries), the research team reviewed the pros and cons of developing a unified console experience. We highlighted both user-centric benefits and quantifiable business impact, forecasting potential efficiency gains while addressing challenges associated with platform consolidation.

Understanding the challenge: Infrastructure tools were hard to discover, different in look and feel, hard to integrate, difficult to onboard to, and time consuming to troubleshoot, we composed a vision document that was shared with key stakeholders for preliminary feedback. We also presented our vision to executive leadership and secured their support by explaining the problem and the impact potential of our proposed solution.

Our presentations were structured to achieve a clear objective: ignite enthusiasm among leadership. Critically, we emphasized that our primary objective was to solicit early feedback on the nascent idea, not to secure immediate resource allocation. This was essential to mitigate the risk of investing substantial effort in an initiative that lacked high-level support.

Early executive involvement ensured alignment and uncovered synergies with ongoing initiatives. Our initial vision for a bespoke console, mirroring Google Cloud Platform's (GCP) console and leveraging its existing infrastructure, was quickly challenged. Because our literature review only focused on human-computer interaction (HCI) costs and benefits, newly discovered technical tradeoffs and limitations associated with replicating GCP made us look for an alternative direction.

We discovered three parallel console initiatives within sister organizations that led to a console evaluation, bringing together PM, Engineering, and UX partners to assess various solutions against a defined set of criteria such as usability, performance, personalization, scalability, security, integration, reusability, maintainability, automation, ML/AI capabilities.

Based on the results of the evaluation we made the choice to adopt a strategy focused on extending an existing internal compute infrastructure management and observability UI (Sigma), to encompass more use cases. Our short-term roadmap (next two years) prioritized optimizing Sigma to better address unmet user needs in key areas such as machine learning and data pipeline jobs.

Our long-term vision (three to five years) maintains our commitment to enhancing Sigma while integrating it with Google's production platform (GPP), an internal platform and console solution that encompasses language libraries, conformance, tests, experimentation, release, rollouts, capacity management, monitoring and incident management.

## 2.4 Vision Implementation

The old, legacy page was too complex and difficult to read. Users reported frustration with the UI's inability to display data in an easy to recall way. To create a shared design language across tools (hypothesis 2), we rolled out a UI redesign that improved readability and significantly reduced the amount of time users needed to find relevant information about their jobs. This change improved readability and made the page look like other popular tools our users used on a daily basis. A scorecard section at the top of the page provided runtime information at a glance, while dedicated sections of different information categories contextually summarized details about users' jobs. Users were also able to personalize these cards and only see information that was relevant to them.

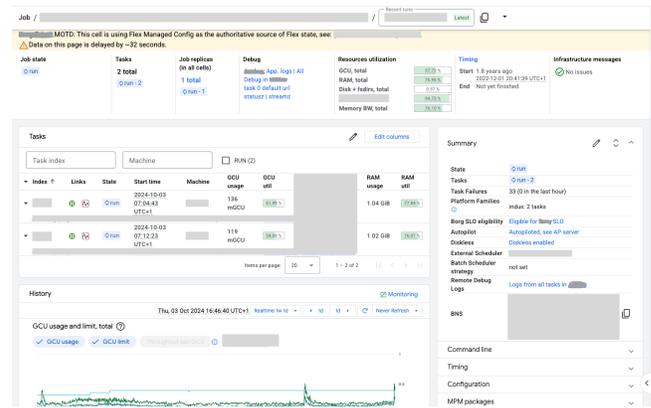

Figure 2: Product UI after the redesign

We also started to take steps towards streamlining the developer experience and consolidating disparate tools (hypothesis 1). We partnered with a team that provides a set of libraries and a language runtime system for developers that can be used to create parallel data-processing pipelines. We defined program goals, assigned responsibilities, set priorities, and crafted a consolidation plan using the RACI model[2].

The UX team designed a high-fidelity prototype (figure 4) and tested it through three rounds of rapid iterative testing and evaluation (RITE) studies. This approach allowed us to better understand users' needs, iterate on the initial concept, and define both the most viable product (MVP) and future goal state of our solution. Feedback from expert and advanced users helped us understand that the most pressing issue was the amount of time it took users to find what was actually wrong with their pipelines. The final design of the dedicated "Pipeline" page features a "Needs Attention" section listing all of the error messages and the relevant warnings that requires users' attention.

Next, our SWE partners will work on implementing the UI changes in the upcoming quarters. We hope that we can release the MVP to a small set of customers by the end of Q2, 2025 and gather feedback from them before giving access to the new features to everyone. Based on telemetry data and feedback from another set of qualitative user interviews we'll identify and fix bugs and usability issues before giving access to the new features to everyone.

## 3 CONCLUSION

This case study illustrates how a focus on user needs can drive positive change within a complex environment like Google. My involvement in this project, from the initial research phase to securing leadership support and implementing incremental improvements, aligns well with the goals of the UXR PoV workshop.

We highlighted our process of creating a strategic vision for tool consolidation and streamlining. We put an emphasis on stakeholder engagement, including gaining executive buy-in, and navigating organizational complexities to achieve meaningful change. The project employed a variety of research methods, including surveys, user interviews, workshops, story mapping, service blueprinting, and R.I.T.E. studies. The TI team also developed a metric based on "SWE hours saved" to demonstrate the return on investment of the work.

I hope that my experience leading this program makes me a good candidate to participate in the UXR POV workshop.

## ACKNOWLEDGMENTS

We extend our gratitude to Amin Vahdat, Peter Backman, and Sumit Gupta for their support and belief in our vision, to John Wilkes and David Hallowell for their invaluable feedback on the case study. Gemini Advanced was utilized to edit parts of the text within this work.